# Searching for Relevant Lessons Learned Using Hybrid Information Retrieval Classifiers: A Case Study in Software Engineering

Tamer Mohamed Abdellatif
Department of Electrical & Computer Engineering
Western University
London, ON, Canada
tmohame7@uwo.ca

Luiz Fernando Capretz
Department of Electrical & Computer Engineering
Western University
London, ON, Canada
lcapretz@uwo.ca

Danny Ho
Department of Electrical & Computer Engineering
Western University
London, ON, Canada
danny@nfa-estimation.com

## ABSTRACT

The lessons learned (LL) repository is one of the most valuable sources of knowledge for a software organization. It can provide distinctive guidance regarding previous working solutions for historical software management problems, or former success stories to be followed. However, the unstructured format of the LL repository makes it difficult to search using general queries, which are manually inputted by project managers (PMs). For this reason, this repository may often be overlooked despite the valuable information it provides. Since the LL repository targets PMs, the search method should be domain specific rather than generic as in the case of general web searching. In previous work, we provided an automatic information retrieval based LL classifier solution. In our solution, we relied on existing project management artifacts in constructing the search query on-the-fly. In this paper, we extend our previous work by examining the impact of the hybridization of multiple LL classifiers, from our previous study, on performance. We employ two of the hybridization techniques from the literature to construct the hybrid classifiers. An industrial dataset of 212 LL records is used for validation. The results show the superiority of the hybrid classifier over the top achieving individual classifier, which reached 25%.

## CCS CONCEPTS

• **Software and its engineering~Software development process management** • Information systems~Retrieval effectiveness

## KEYWORDS

Professional search, project management lessons learned, lessons learned recall, hybrid classifiers, information retrieval

## 1 INTRODUCTION AND BACKGROUND

Software organizations supposedly store their historical data in lessons learned (LL) repositories. This data can be success stories or solutions to issues that were discovered in previous projects, which can be reused in similar future situations. On the other hand, this data can also include failure stories, pitfalls or mistakes from previous projects to be avoided in similar future projects. Accordingly, the LL repositories contain information that can be useful in guiding project managers (PMs) to leverage opportunities or avoid repeating past mistakes. For example, an LL record concerning a decision about whether to implement a mobile application in-house or outsource the implementation can take the following form:

*Context*: the project scope includes an implementation of a small-sized mobile application. This mobile application is not reusable, i.e., it will only be used in this project.

*Problem*: if the mobile application is of a small size, then the organizational process overhead, such as quality assurance and management reporting tasks, will affect the profit of implementing the mobile application in-house.

*Recommended actions*: outsource the implementation to an external mobile application specialized company. Contact the purchase team for a trusted partners list.

However, this LL information can only be beneficial if project managers refer to it frequently to solve present issues or to seek potential opportunities, which is not always the case. Unfortunately, LL are often abandoned by PMs due to the effort and time needed to manually search for relevant LL records within the unstructured, i.e., natural language, LL repository [11]. Also, it has been found to be difficult to search for relevant LL records using a general search methodology or search terms manually defined by PMs. This calls for automatic domain specific, i.e., professional search, LL recall solutions [12]. By *automatic* we mean that there should be no need for manual searching to facilitate the exploitation of LL. In a previous study, we worked on satisfying this need by providing an automatic domain specific LL recall, i.e., retrieval, system [1].

Regarding the software engineering literature, there is a paucity of software engineering research addressing LL recall solutions [12]. As per our knowledge, the most relevant studies have been conducted by Sary and Mackey [9] and by Weber et al. [13]. Both studies employed case-based reasoning techniques in order to build their systems. Also, these studies have the common limitation of the need to arrange the LL repository records in a question-answer format. This transformation is difficult to achieve in reality as it demands extra effort and time. This limitation is not valid in our solution since the classifier is constructed using the LL repository records *as is*, with no transformation needed. Also, these studies are different from our solution since we employ IR techniques instead of case-based reasoning. As per our knowledge, we are the first to apply the IR models to the software project management LL recall context [4].

In order to make our solution specific to the project management LL domain, we relied on two of the existing and most influential project management artifacts, namely project management issues and risks, to automatically invoke our constructed classifiers. Since these artifacts are already associated with the software development project lifecycle, there is no need for the manual involvement of PMs. We relied on some of the most popular information retrieval (IR) models from the literature to construct the LL classifiers. In addition, we evaluated our solution through an empirical case study using a real dataset from industry [1]. The results of the case study proved the effectiveness of our solution as it achieved about 70% accuracy.

In this paper, we conduct an extension case study. In this study, we constructed hybrid LL classifiers by combining multiple LL classifiers from our previous work in order to examine the impact of this hybridization on the performance of the LL classifiers. Our main motive for conducting such an extension is that, although several domains have studied classifiers' hybridization [6,10], it has not been studied in the LL recall context.

The results from our extension study showed an improvement in the majority of the hybrid cases that were studied. Although some of the cases showed no improvement or showed a decrease in the performance of the hybrid classifiers, the fact that the hybridization was successful in most of the other cases, and in other software engineering domains [6,10], makes considering the hybrid classifiers interesting for future studies regarding LL recall.

## 2 CASE STUDY DESIGN

### 2.1 Previous Case Study Summary

In our previous work [1], we provided a solution to improve the retrieval of the software LL and make them available to PMs. We relied on two of the software project management artifacts, namely project management issues and risk registers. These two artifacts were used to construct a query string on-the-fly. The constructed query string was then used to *automatically* call an LL classifier in order to retrieve LL records relevant to the project at hand. Regarding the LL classifiers, we employed three of the popular IR models to construct the classifiers, and we considered multiple parameter values to configure and construct multiple classifiers. The employed IR models were the Vector Space Model (VSM) [7], the Latent Semantic Indexing (LSI) [5], and the Latent Dirichlet Allocation (LDA) [7,8]. The parameter values considered for the VSM were as follows: *term weight* (tf-idf, sublinear tf-idf, boolean), where tf is term frequency and idf is inverse document frequency, and *similarity* (cosine, overlap) [6,8]. The parameter values for LSI were *term weight* (tf-idf, sublinear tf-idf, boolean), *similarity* (cosine), and *number of topics* (32, 64, 128, 256). For LDA, the parameter values were *number of topics* (32, 64, 128, 256) and *similarity* (conditional probability).

In order to reduce the noise in the input data, we considered two of the preprocessing steps from the natural language processing literature, namely *stemming* and *stopping* steps [8]. In the stemming step, the word is replaced by its morphological root. In the stopping step, commonly used words, such as "the" in the English language, are removed [8]. We considered studying the constructed classifier by applying four combinations of these steps to the input data: 1) none of the preprocessing steps were applied, 2) only the stemming step was applied, 3) only the stopping step was applied, and 4) both steps were applied together.

In our previous case study, we considered all combinations for all IR models, parameter values and input preprocessing steps, which led to the construction of 88 LL classifiers. The performance of all the considered classifiers was validated using the top-K performance metric from the IR literature [6,8]. Top-K, top-20 in our study, examines the number of queries where the classifier returns at least one relevant record within the first K items of the retrieved list.

Also, for our solution validation, we relied on a real industrial dataset provided by a multinational software development partner. The validation dataset included 212 real LL records from 30 projects and 55 project management issues and risk records. In addition, both the performance results and the impact of each parameter value on the results were statistically analyzed. A satisfactory maximum performance result of 70% for the top-20 was recorded, which positively proved the effectiveness of our provided solution [1].

## 2.2 Lessons Learned Hybrid Classifiers

Different classifiers can perform in different ways regarding the same dataset and inputs. This means that different classifiers can have different errors and advantages. Thus, combining multiple classifiers together can either optimistically lead to better performance as they complement each other to avoid individual errors, or negatively lead to worse performance when they distract each other. This depends heavily on the chosen classifiers. Based on this fact, we aim, in our case study, to combine multiple classifiers from our previous work to construct a hybrid classifier, and then study the impact of this combination on performance.

*2.2.1 Hybridization Techniques.* We employed two hybridization techniques from the literature [10] to combine individual classifiers into one hybrid classifier. These two techniques are *Borda count* and *Score Addition*.

The *Borda* technique is a rank-based technique. This means that it relies on the rank, i.e., the order in the retrieved list, of the retrieved item, the relevant LL in our case, within the classification results list from each individual classifier, to assign this item a rank score. The *Borda* count can be calculated as stated in [10] as

$$Borda\ (d_k) = \sum_{C_i \in C} M_i - r(d_k \mid C_i) + 1 , \qquad [10]$$

where $d_k$ is the retrieved list item for which the Borda count is calculated, $C$ is the collection of the hybrid classifiers, $C_i$ is the ith classifier within the $C$ collection, $M_i$ is the number of retrieved items that received a non-zero score in the retrieved list by the classifier $C_i$, and $r(d_k \mid C_i)$ is the $d_k$ rank or order within the $C_i$ retrieved list [10].

On the other hand, the score addition technique relies on the item's weight or the score given by the individual classifiers. The total hybrid score of each retrieved item is calculated as the summation of the individual score of this item from each of the combined classifiers [10]. In order to avoid any mistaken bias to a certain classifier due to the weighing scale, the items' weights in each of the combined classifiers list are scaled to be within the same range of [0-1]. Accordingly, the individual item's score addition can be calculated as follows:

$$ScoreAddition(d_k) = \sum_{C_i \in C} s(d_k \mid C_i) , \qquad [10]$$

where $s(d_k \mid C_i)$ is the score of $d_k$ given by the classifier $C_i$ [10]. Finally, the items are ordered descendingly, based on their total score.

*2.2.2 Hybrid Classifiers Selection.* The selection of the combined classifiers has a crucial impact on the performance of the constructed hybrid classifier. For this reason, we aimed to choose the classifiers which can positively complement each other. Thus, we chose the classifiers which were exposed to different formats of the input data, because such classifiers could have higher chances of coming up with different insights and conclusions regarding the dataset at hand, which we thought could improve their combined performance. That said, we decided to proceed with the classifiers which were constructed by applying different input preprocessing step combinations. These preprocessing steps were employed in four combinations, as described in Section 2.1, leading to four classifier subspaces. So, for each IR model, we considered a top performer classifier from each of the four classifier subspaces. This resulted in the selection of four classifiers from each of the VSM, LSI, and LDA models that included: the top classifier when none of the preprocessing steps were applied, the top classifier when the stemming step was applied, the top classifier when the stopping step was applied, and finally the top performer classifier when both the stemming and stopping steps were applied together. In our experiment, we examined the performance of the hybrid classifiers constructed by combining the four selected classifiers of each IR model in pairs. In addition to studying these pairs of classifier combinations, we

**Table 1 Hybrid Classifiers Results**

| | Top-20 Results | | | | | | Top-20 Results | | | | |
|---|---|---|---|---|---|---|---|---|---|---|---|
| Comb. ID | Top Individual Performance (%) | Score Addition (%) | RI (%) | Borda Count (%) | RI (%) | Comb. ID | Top Individual Performance (%) | Score Addition (%) | RI (%) | Borda Count (%) | RI (%) |
| 1 | 46 | 50 | 8 | 56 | 20 | 12 | 70 | 74 | 5 | 69 | -3 |
| 2 | 46 | 52 | 12 | 57 | 24 | 13 | 69 | 69 | 0 | 69 | 0 |
| 3 | 52 | 50 | -4 | 50 | -4 | 14 | 70 | 70 | 0 | 70 | 0 |
| 4 | 52 | 54 | 4 | 56 | 7 | 15 | 61 | 61 | 0 | 59 | -3 |
| 5 | 52 | 46 | -11 | 44 | -14 | 16 | 61 | 70 | 15 | 65 | 6 |
| 6 | 41 | 46 | 14 | 44 | 9 | 17 | 61 | 61 | 0 | 59 | -3 |
| 7 | 52 | 48 | -7 | 48 | -7 | 18 | 70 | 65 | -8 | 63 | -11 |
| 8 | 69 | 69 | 0 | 70 | 3 | 19 | 70 | 70 | 0 | 70 | 0 |
| 9 | 69 | 70 | 3 | 72 | 5 | 20 | 70 | 72 | 3 | 72 | 3 |
| 10 | 70 | 67 | -5 | 70 | 0 | 21 | 70 | 70 | 0 | 65 | -8 |
| 11 | 70 | 72 | 3 | 69 | -3 | 22 | 70 | 72 | 3 | 72 | 3 |

Comb. ID = combination id

studied the performance of the combination of the four selected classifiers in each IR model as well. Finally, we combined all of the selected classifiers from all IR models together (four classifiers from each of the three IR models considered). All the classifier combinations are shared in detail online for reference [2].

## 3 CASE STUDY RESULTS AND DISCUSSION

The performance results for each of the constructed hybrid classifiers were compared to the performance results of each of the combined classifiers using the relative performance improvement (RI) percentage. The RI calculation is formulated as follows:

$$RI = \frac{P(HC) - HighestP\ (Combined\ Classifiers)}{HighestP\ (Combined\ Classifiers)},$$

where $P(HC)$ is the value of the performance metric $P$ for the hybrid classifier $HC$, and the $HighestP()$ method returns the highest performance metric value among the performance values of the combined classifiers [10].

The results for the considered hybrid classifiers and the impact on the top-20 are shown in Table 1. In the case of using the score addition method, the hybrid classifier results showed either an improvement or no effect against the individual classifiers in about 77% of the cases considered. In other words, the score addition combination led to a decrease in performance in only five cases. Regarding the Borda method, there was an improvement or no effect in about 59% of the cases. The maximum improvement was 15% for the score addition method and 24% for the Borda method.

An additional important observation is that the combination performance exceeded the 70.37% top-20, which was the top performance recorded among all the individual classifiers in our previous experimental work. For score addition, this was recorded in four cases where top-20 performance accuracies of 74.07% and 72.22% were recorded. In the case of Borda, this was achieved in three cases where a top-20 of 72.22% was recorded. Also, it is important to highlight that in approximately 73% of the cases, the score addition results outperformed or were comparable to the Borda results.

Although the hybridization did not prove to be an improvement in all cases within our experiment, the number of the improved cases, especially the 77% of cases for score addition, are considered satisfactory and encourage the consideration of hybrid classifiers within the LL retrieval context.

## 4 CONCLUSIONS

In this paper, we provided an extension of our previous empirical study regarding the construction of an automatic software management LL recall system. Our solution represented a domain specific search, i.e., professional search, as we constructed the search query using two of the existing project management artifacts instead of employing a generic manual search. In our extension, we studied the impact of the hybridization of the LL classifiers on performance. We relied on the existing LL classifiers from our previous study in constructing the hybrid classifiers. In this study, we employed two combination techniques in constructing the hybrid classifiers. A comparison was conducted between the performance of each hybrid classifier and the performance of the top performer from the combined individual classifiers. The top-K performance metric was employed to measure the retrieval accuracy of the classifiers considered. The study results showed a relative improvement, or no effect, of the hybrid classifiers' performance against the individual classifiers' performance in about 77% of the cases in the top-20 using the score addition method. Although, the improvement was not satisfactory in some cases, the overall results were encouraging and provided positive insights regarding employing IR models to provide a domain specific LL recall solution.

## REFERENCES

[1] Author1, Author2, and Author3. 2018. Automatic Recall of Software Lessons Learned for Software Project Managers. (Submitted)
[2] Author1. 2018. Top-K Hybrid Results: https://drive.google.com/file/d/1IbfkQtJ-fMbYuQU5eh2jGxSQpS8fgXAr/view?usp=sharing
[3] D. M. Blei, A. Y. Ng, and M. I. Jordan. 2003. Latent Dirichlet Allocation. *Journal of Machine Learning Research* 3 (Jan 2003). 993–1022.
[4] T. H. Chen, S. W. Thomas, and A. E. Hassan. 2016. A Survey on the Use of Topic Models When Mining Software Repositories. *Empirical Software Eng.* 21, 5 (2016), 1843–1919.
[5] S. Deerwester, S. T. Dumais, G. W. Furnas, T. K. Landauer, and R. Harshman. 1990. Indexing by Latent Semantic Analysis. *Journal of the American Society for Information Science* 41, 6 (1990), 391–407.
[6] E. Kocaguneli, T. Menzies, and J. W. Keung. 2012. On the Value of Ensemble Effort Estimation. *IEEE Trans. Software Eng.* 38, 6 (Nov./Dec. 2012), 1403–1416.
[7] C. D. Manning, P. Raghavan, and H. Schutze. 2008. *Introduction to Information Retrieval*, Vol. 1, Cambridge University Press, Cambridge.
[8] T. Mens, A. Serebrenik, and A. Cleve (Eds.). 2014. *Evolving Software Systems*. Springer-Verlag, Berlin Heidelberg, Chapter 5.
[9] C. Sary and W. Mackey. 1995. A Case-based Reasoning Approach for the Access and Reuse of Lessons Learned. *In Proceeding of the 5th Annual International Symposium of National Council on Systems Engineering*. St. Louis, Missouri, 249–256.
[10] S. W. Thomas, M. Nagappan, D. Blostein, and A. E. Hassan. 2013. The Impact of Classifier Configuration and Classifier Combination on Bug Localization. *IEEE Trans. Software Eng.* 39, 10 (Oct. 2013), 1427–1443.
[11] R. Weber and D. Aha. 2002. Intelligent Delivery of Military Lessons Learned. *Decision Support Systems* 34, 3 (2002), 287–304.
[12] R. Weber, D. W. Aha, and I. Becerra-Fernandez. (2001). Intelligent Lessons Learned Systems. *Expert Systems with Applications* 20, 1 (2001), 17–34.
[13] R. Weber, D. W. Aha, L. K. Branting, J. R. Lucas, and I. Fernandez. 2000. Active Case-Based Reasoning for Lessons Delivery Systems. *In Proceeding of the 13th Annual Conference of the International Florida Artificial Intelligence Research Society* (FLAIRS'00*)*, AAAI Press, Orlando, FL, 170–174.